\documentclass{revtex4}
\usepackage{amssymb}
\usepackage{amsmath}

\input{tcilatex}

\begin{document}

\title{Existence of Reissner-Nordstr\"{o}m type black holes in f(R) gravity }
\author{S. Habib Mazharimousavi}
\email{habib.mazhari@emu.edu.tr}
\author{M. Kerachian}
\email{kerachian.morteza@gmail.com}
\author{M. Halilsoy}
\email{mustafa.halilsoy@emu.edu.tr}
\affiliation{Physics Department, Eastern Mediterranean University, G. Magusa north
Cyprus, Mersin 10 Turkey. }

\begin{abstract}
We investigate the existence of Reissner-Nordstr\"{o}m (RN) type black holes
in f(R) gravity. Our emphasis is to derive, in the presence of electrostatic
source, the necessary conditions which provide such static, spherically
symmetric (SSS) black holes available in f(R) gravity. We also study the
thermodynamics of the black hole solution.


\end{abstract}

\maketitle

\section{Introduction}

Due to a number of valid reasons $f(R)$ gravity attracted much interest
during the recent decade as an extension / modification of Einstein's
general relativity \cite{1,2,3,4,5,6} (for some review works see \cite%
{7,8,9,10}). Here $R$ stands for the Ricci scalar, the simplest among\ much
complicated ones and $f(R)$ is an analytic function of $R$. Herein we wish
to look at $f(R)$ gravity from a different angle\ which was introduced by
Bergliaffa and Nunes in their novel paper \cite{11,12}. Since the black hole
solutions in Einstein's $f(R)=R$, theory has already built enough prominence
and play the leading role it should be wise to seek\ for similar solutions
in the more general $f(R)$ theories. This approach concerns directly the
existence problem of black holes and it's associated necessary conditions
for analog objects in the latter. The existence conditions may simply be
dubbed\ as the "\textit{near-horizon test}" in order to highlight the event
horizon of a black hole as a physical reality. It is well-known that
physically when the observer approaches the event horizon he / she feels
nothing unusual except strong gravity, so this mathematically must reflect
analytically on the event horizon. The analytic expansion of a metric
function, say $f(r)$, is developed\ in series of the form $%
f(r)=f(r_{0})+f^{\prime }(r_{0})\left( r-r_{0}\right) +\mathcal{O}\left(
\left( r-r_{0}\right) ^{2}\right) $ where $r_{0}$ is the event horizon and $%
\left( r-r_{0}\right) $ stands naturally small. When these developed series
are substituted back into the Einstein equations they will give conditions
of zeroth,first and higher orders. These are precisely what we call the
necessary conditions for the existence of certain / analog black hole types.
To our amazement\ these necessary conditions emerge rather restrictive so
that we can't propose arbitrarily any polynomial forms of $f(R)$ as the
representative black holes. For instance,what are the necessary conditions
in order that it will admit Schwarzschild-like black hole solutions? This
particular problem without external sources has already been considered and
it's found that the possible\ $f(R)$ must be of the form $f(R)=\alpha \sqrt{%
R+\beta }$, in which $\alpha $ and $\beta $ are constants \cite{11,12}. An
analytic expansion reveals that the first term retains the Einstein-Hillbert
term with the addition of higher orders in $R$ which seems to be the payoff
in the enterprise of $f(R)$ theory. Any $f(R)$ \ theory is known to create
it's own source\ from the inherent\ non-linearity of the theory. Beside
these, however, additional external sources\ may be considered (some
examples of $f(R)$ black hole with charge are given in \cite%
{13,14,15,16,17,18,19}),\ which makes the principal aim of the present\
paper. We consider an external static electric field as source and adopt the
Reissner-Nordstr\"{o}m (RN)-type black hole within $f(R)$ gravity.
Expectedly, the results for necessary conditions for the existence of a RN
black hole are more complicated than the case of a\ Schwarzschild black
hole. In this process we obtain an infinite series representation for the
near-horizon behavior of our metric functions.\ The exact determination of
the constant coefficients in the series is theoretically possible, at least\
in the leading orders. The addition of further external sources beside
electromagnetism\ will naturally make the problem more complicated. An
equally simple case is the extremal RN black hole which is also considered
in our study.

The paper is organized as follows. Section II investigates the necessary
conditions for the existence of a RN-type black hole in $f(R)$ gravity.
Thermodynamics, and in particular the first law for such black holes are
presented in Section III. Section IV is devoted to an extremal RN-type black
hole. The paper is completed with Concluding Remarks, which appears in
Section V.

\section{Analog RN black holes in $f(R)$ gravity}

The proper action in $f(R)$ gravity coupled minimally with Maxwell source in 
$4-$dimensions is given by%
\begin{equation}
S=\dint \sqrt{-g}\left( \frac{f(R)}{2\kappa }-\frac{\mathcal{F}}{4\pi }%
\right) d^{4}x
\end{equation}%
in which $f(R)$ is a real function of the Ricci scalar $R,$ $\mathcal{F}=%
\frac{1}{4}F_{\mu \nu }F^{\mu \nu }$ is the Maxwell invariant and $\kappa
=8\pi G$ where $G$ is the Newton's constant. Our choice of the spacetime is
a RN-type black hole solution whose line element can be written as%
\begin{equation}
ds^{2}=-e^{-2\Phi }\left( 1-\frac{2M}{r}+\frac{Q^{2}}{r^{2}}\right) dt^{2}+%
\frac{dr^{2}}{\left( 1-\frac{2M}{r}+\frac{Q^{2}}{r^{2}}\right) }+r^{2}\left(
d\theta ^{2}+\sin ^{2}\theta d\varphi ^{2}\right) ,
\end{equation}%
where $M$ and $Q$ are two real constants which indicate the mass and the
charge of the black hole respectively. Also $\Phi =\Phi \left( r\right) $ is
an unknown real function which is well behaved everywhere and dies off at
large $r.$ The matter source which we shall consider in our consideration is
a Maxwell electric field whose two-forms is given by

\begin{equation}
\mathbf{F}=E\left( r\right) dt\wedge dr
\end{equation}%
where $E\left( r\right) $ is the electric field. Following the line element
(2) one finds the dual-Maxwell field as 
\begin{equation}
^{\ast }\mathbf{F}=-E\left( r\right) e^{\Phi }r^{2}\sin \theta d\theta
\wedge d\varphi
\end{equation}%
and in turn the Maxwell equation%
\begin{equation}
d^{\ast }\mathbf{F}=0
\end{equation}%
implies 
\begin{equation}
E\left( r\right) =\frac{q}{r^{2}}e^{-\Phi }
\end{equation}%
in which the integration constant $q$ is to be identified with $Q$. Varying
the action with respect to $g_{\mu \nu }$ provides the field equations%
\begin{equation}
FR_{\mu \nu }-\frac{1}{2}fg_{\mu \nu }-\nabla _{\mu }\nabla _{\nu }F+g_{\mu
\nu }\square F=\kappa T_{\mu \nu }
\end{equation}%
where $F=\frac{df}{dR}$ \ , $\square F=\frac{1}{\sqrt{-g}}\partial _{\mu
}\left( \sqrt{-g}\partial ^{\mu }\left( \frac{df}{dR}\right) \right) $ \ \
and \ $\nabla ^{\nu }\nabla _{\mu }F=g^{\alpha \nu }\left[ F_{,\mu ,\alpha
}-\Gamma _{\mu \alpha }^{m}F_{,m}\right] .$ We obtain

\begin{equation}
\square F=\square \frac{df}{dR}=\frac{1}{\sqrt{-g}}\partial _{r}\left( \sqrt{%
-g}g^{rr}\partial _{r}F\right) ,
\end{equation}

\begin{equation}
\nabla ^{t}\nabla _{t}F=\frac{1}{2}g^{tt}g^{rr}g_{tt,r}F^{\prime },
\end{equation}

\begin{equation}
\nabla ^{r}\nabla _{r}F=g^{rr}F^{\prime \prime }-g^{rr}\Gamma
_{rr}^{r}F^{\prime },
\end{equation}

\begin{equation}
\nabla ^{\varphi }\nabla _{\varphi }F=\nabla ^{\theta }\nabla _{\theta }F=%
\frac{1}{2}g^{\theta \theta }g^{rr}g_{\theta \theta ,r}F^{\prime }
\end{equation}%
in which a prime denotes derivative with respect to $r.$ Also in Eq. (7) the
stress-energy tensor $T_{\mu }^{\nu }$ reads as 
\begin{equation}
T_{\mu }^{\nu }=-\frac{1}{4\pi }\left( \mathcal{F}\delta _{\mu }^{\nu
}-F_{\mu \lambda }F^{\nu \lambda }\right) ,
\end{equation}%
which after considering the line element (2) and the Maxwell field (3)
together with (6), one finds 
\begin{equation}
T_{\mu }^{\nu }=\frac{1}{8\pi }\frac{Q^{2}}{r^{4}}\text{diag}\left[ -1,-1,1,1%
\right] .
\end{equation}%
We note that another dependent equation is the vanishing trace condition%
\begin{equation}
FR-2f+3\square F=0,
\end{equation}%
which is obtained after knowing $T=T_{\mu }^{\mu }=0$. The trace equation
may be used to simplify the field equations and therefore Eq. (7) becomes 
\begin{equation}
FR_{\nu }^{\mu }-\frac{1}{4}\delta _{\nu }^{\mu }\left( FR-\square F\right)
-\nabla ^{\mu }\nabla _{\nu }F=\kappa T_{\nu }^{\mu }.
\end{equation}%
From the metric given in (2) one finds the event horizon at $r=r_{0}=M+\sqrt{%
M^{2}-Q^{2}}$ or consequently%
\begin{equation}
M=\frac{r_{0}^{2}+Q^{2}}{2r_{0}},
\end{equation}%
as the ADM mass. Based on the \textit{near horizon test} introduced in Ref. 
\cite{11,12} we expand all the unknown functions about the horizon. This
would lead to the expansions%
\begin{equation}
R\left( r\right) =R_{0}+R_{0}^{\prime }\left( r-r_{0}\right) +\frac{1}{2}%
R_{0}^{\prime \prime }\left( r-r_{0}\right) ^{2}+\mathcal{O}\left( \left(
r-r_{0}\right) ^{3}\right) ,
\end{equation}%
\begin{equation}
\Phi \left( r\right) =\Phi _{0}+\Phi _{0}^{\prime }\left( r-r_{0}\right) +%
\frac{1}{2}\Phi _{0}^{\prime \prime }\left( r-r_{0}\right) ^{2}+\mathcal{O}%
\left( \left( r-r_{0}\right) ^{3}\right) ,
\end{equation}%
\begin{equation}
F=F_{0}+F_{0}^{\prime }\left( r-r_{0}\right) +\frac{1}{2}F_{0}^{\prime
\prime }\left( r-r_{0}\right) ^{2}+\mathcal{O}\left( \left( r-r_{0}\right)
^{3}\right) ,
\end{equation}%
in which the sub zero implies the corresponding quantity evaluated at the
horizon. After some manipulation, the field equations would develop as
series in different orders of $\left( r-r_{0}\right) .$ In the zeroth order
one finds two independent equations

\begin{equation}
f_{0}r_{0}^{4}-\left( E_{0}R_{0}^{\prime }+3\Phi _{0}^{\prime }F_{0}\right)
r_{0}^{3}+Q^{2}\left( E_{0}R_{0}^{\prime }+3\Phi _{0}^{\prime }F_{0}\right)
r_{0}+2Q^{2}\left( F_{0}-1\right) =0,
\end{equation}%
\begin{equation}
f_{0}r_{0}^{4}-2r_{0}^{3}E_{0}R_{0}^{\prime }+2Q^{2}r_{0}E_{0}R_{0}^{\prime
}-2Q^{2}\left( F_{0}-1\right) =0,
\end{equation}%
together with 
\begin{equation}
R_{0}=\frac{3\Phi _{0}^{\prime }\left( r_{0}^{2}-Q^{2}\right) }{r_{0}^{3}}.
\end{equation}%
The first order equations admit another pair of equations 
\begin{gather}
F_{0}R_{0}^{\prime }r_{0}^{4}+\left[ \left( 2\Phi _{0}^{\prime 2}-5\Phi
_{0}^{\prime \prime }\right) F_{0}-3\Phi _{0}^{\prime }E_{0}R_{0}^{\prime
}-3H_{0}R_{0}^{\prime 2}+4f_{0}-3E_{0}R_{0}^{\prime \prime }\right]
r_{0}^{3}-2\left( 3E_{0}R_{0}^{\prime }+5\Phi _{0}^{\prime }F_{0}\right) + \\
\left[ \left( -2\Phi _{0}^{\prime 2}+5\Phi _{0}^{\prime \prime }\right)
F_{0}+3\Phi _{0}^{\prime }E_{0}R_{0}^{\prime }+3H_{0}R_{0}^{\prime
2}+3E_{0}R_{0}^{\prime \prime }\right] Q^{2}r_{0}+6Q^{2}E_{0}R_{0}^{\prime
}+4\Phi _{0}^{\prime }F_{0}Q^{2}=0,  \notag
\end{gather}%
\begin{gather}
F_{0}R_{0}^{\prime }r_{0}^{4}+4\left( f_{0}-E_{0}R_{0}^{\prime \prime
}-H_{0}R_{0}^{\prime 2}+\frac{1}{2}\Phi _{0}^{\prime }E_{0}R_{0}^{\prime
}\right) r_{0}^{3}-2\left( \Phi _{0}^{\prime }F_{0}+3E_{0}R_{0}^{\prime
}\right) r_{0}^{2}+ \\
\left( 4E_{0}R_{0}^{\prime \prime }-2\Phi _{0}^{\prime }E_{0}R_{0}^{\prime
}+4H_{0}R_{0}^{\prime 2}\right) Q^{2}r_{0}+2\Phi _{0}^{\prime }F_{0}Q^{2}=0 
\notag
\end{gather}%
together with%
\begin{equation}
R_{0}^{\prime }=\frac{\left( 5\Phi _{0}^{\prime \prime }-2\Phi _{0}^{\prime
2}\right) r_{0}^{3}-2\Phi _{0}^{\prime }r_{0}^{2}+\left( 2Q^{2}\Phi
_{0}^{\prime 2}-5Q^{2}\Phi _{0}^{\prime \prime }\right) r_{0}+8Q^{2}\Phi
_{0}^{\prime }}{r_{0}^{4}}.
\end{equation}%
In these equations $E=\frac{d^{2}f}{dR^{2}}=\frac{dF}{dR}$ and $H=\frac{%
d^{3}f}{dR^{3}}=\frac{dE}{dR}$ and a sub "$0$" implies the value at the
horizon. From these equations we find the possible solutions for the unknown
coefficients. The following are the results:%
\begin{equation}
\Phi =\beta _{1}\epsilon +\beta _{2}\epsilon ^{2}+O\left( \epsilon
^{3}\right)
\end{equation}%
\begin{equation}
f=f_{0}-\frac{1}{6}\frac{\left( f_{0}r_{0}^{4}-6Q^{2}\right) \left[
2r_{0}\left( r_{0}^{2}-Q^{2}\right) \beta _{1}^{2}+2\left(
r_{0}^{2}-4Q^{2}\right) \beta _{1}-5r_{0}\left( r_{0}^{2}-Q^{2}\right) \beta
_{2}\right] }{r_{0}^{4}\left( r_{0}\left( r_{0}^{2}-Q^{2}\right) \beta
_{1}-Q^{2}\right) }\epsilon +O\left( \epsilon ^{2}\right)
\end{equation}%
\begin{equation}
F=\frac{f_{0}r_{0}^{4}-6Q^{2}}{6\left( \beta _{1}r_{0}\left(
r_{0}^{2}-Q^{2}\right) -Q^{2}\right) }+\frac{3\beta _{1}\left(
r_{0}^{2}-Q^{2}\right) \left( r_{0}^{4}f_{0}+2Q^{2}\right)
-4f_{0}r_{0}^{3}Q^{2}}{6\left( r_{0}^{2}-Q^{2}\right) \left( \beta
_{1}r_{0}\left( r_{0}^{2}-Q^{2}\right) -Q^{2}\right) }\epsilon +O\left(
\epsilon ^{2}\right) ,
\end{equation}%
\begin{equation}
R=\frac{3\beta _{1}\left( r_{0}^{2}-Q^{2}\right) }{r_{0}^{3}}-\frac{%
2r_{0}\left( r_{0}^{2}-Q^{2}\right) \beta _{1}^{2}+2\left(
r_{0}^{2}-4Q^{2}\right) \beta _{1}-5r_{0}\left( r_{0}^{2}-Q^{2}\right) \beta
_{2}}{r_{0}^{4}}\epsilon +O\left( \epsilon ^{2}\right)
\end{equation}%
in which $\epsilon =r-r_{0}$, $\beta _{1}$, $\beta _{2}$ are constants and 
\begin{equation}
f_{0}=-\frac{6Q^{2}}{r_{0}^{3}}\frac{8r_{0}\left( r_{0}^{2}-Q^{2}\right)
^{2}\beta _{1}^{2}-2\left( r_{0}^{2}-Q^{2}\right) \left(
Q^{2}+5r_{0}^{2}\right) \beta _{1}-5r_{0}\left( r_{0}^{2}-Q^{2}\right)
^{2}\beta _{2}}{16r_{0}^{2}\left( r_{0}^{2}-Q^{2}\right) ^{2}\beta
_{1}^{2}+2r_{0}\left( r_{0}^{2}-Q^{2}\right) \left(
5r_{0}^{2}-23Q^{2}\right) \beta _{1}+5r_{0}^{2}\left( r_{0}^{2}-Q^{2}\right)
^{2}\beta _{2}+24Q^{4}}.
\end{equation}%
Let us note that $\Phi _{0}$ remains unknown, but since it can be absorbed
into the redefinition of time it can be set as $\Phi _{0}=0.$ What we have
here are some complicated relations between the forms of $f,F$ and $R$ in
terms of $\beta _{1}$ and $\beta _{2}$ which are arbitrary. In the zeroth
order the conditions on any $f(R)$ can be written as 
\begin{equation}
\left. f\right\vert _{r_{0}}=-\frac{6Q^{2}}{r_{0}^{3}}\frac{8r_{0}\left(
r_{0}^{2}-Q^{2}\right) ^{2}\beta _{1}^{2}-2\left( r_{0}^{2}-Q^{2}\right)
\left( Q^{2}+5r_{0}^{2}\right) \beta _{1}-5r_{0}\left(
r_{0}^{2}-Q^{2}\right) ^{2}\beta _{2}}{16r_{0}^{2}\left(
r_{0}^{2}-Q^{2}\right) ^{2}\beta _{1}^{2}+2r_{0}\left(
r_{0}^{2}-Q^{2}\right) \left( 5r_{0}^{2}-23Q^{2}\right) \beta
_{1}+5r_{0}^{2}\left( r_{0}^{2}-Q^{2}\right) ^{2}\beta _{2}+24Q^{4}}
\end{equation}%
and%
\begin{equation}
\left. F\right\vert _{r_{0}}=\frac{f_{0}r_{0}^{4}-6Q^{2}}{6\left( \beta
_{1}r_{0}\left( r_{0}^{2}-Q^{2}\right) -Q^{2}\right) }.
\end{equation}

\subsection{Examples of $f(R)=R$ and $f(R)=R^{2}$}

For instance, in the case of $R$ gravity we have $f_{0}=R_{0}$ and $F=1.$
The latter yields%
\begin{equation}
\frac{\left( \beta _{1}\left( r_{0}^{2}-Q^{2}\right) \right) r_{0}-2Q^{2}}{%
2\left( \beta _{1}r_{0}\left( r_{0}^{2}-Q^{2}\right) -Q^{2}\right) }=1
\end{equation}%
or consequently $\beta _{1}=0.$ Having $\beta _{1}=0$ implies%
\begin{equation}
f_{0}=R_{0}\rightarrow -\frac{6Q^{2}}{r_{0}^{3}}\frac{-5r_{0}\left(
r_{0}^{2}-Q^{2}\right) ^{2}\beta _{2}}{5r_{0}^{2}\left(
r_{0}^{2}-Q^{2}\right) ^{2}\beta _{2}+24Q^{4}}=0
\end{equation}%
which clearly leads to $\beta _{2}=0$. Therefore $f(R)=R$ satisfies our
general conditions with $\beta _{1}=0=\beta _{2}.$ Next we test the case of $%
f(R)=R^{2}$ for which the above conditions become%
\begin{equation}
R_{0}^{2}=-\frac{6Q^{2}}{r_{0}^{3}}\frac{8r_{0}\left( r_{0}^{2}-Q^{2}\right)
^{2}\beta _{1}^{2}-2\left( r_{0}^{2}-Q^{2}\right) \left(
Q^{2}+5r_{0}^{2}\right) \beta _{1}-5r_{0}\left( r_{0}^{2}-Q^{2}\right)
^{2}\beta _{2}}{16r_{0}^{2}\left( r_{0}^{2}-Q^{2}\right) ^{2}\beta
_{1}^{2}+2r_{0}\left( r_{0}^{2}-Q^{2}\right) \left(
5r_{0}^{2}-23Q^{2}\right) \beta _{1}+5r_{0}^{2}\left( r_{0}^{2}-Q^{2}\right)
^{2}\beta _{2}+24Q^{4}}
\end{equation}%
and 
\begin{equation}
2R_{0}=\frac{R_{0}^{2}r_{0}^{4}-6Q^{2}}{6\left( \beta _{1}r_{0}\left(
r_{0}^{2}-Q^{2}\right) -Q^{2}\right) }.
\end{equation}%
These two conditions admit%
\begin{equation}
\beta _{1}=\frac{1}{6}\frac{4Q+2\sqrt{4Q^{2}-2r_{0}^{4}}}{r_{0}\left(
r_{0}^{2}-Q^{2}\right) }
\end{equation}%
and 
\begin{gather}
\beta _{2}= \\
\frac{-4Q\left\{ \left[ Q^{2}\left( 20Q^{2}-11r_{0}^{4}\right)
+15r_{0}^{2}\left( r_{0}^{4}-2Q^{2}\right) \right] \sqrt{4Q^{2}-2r_{0}^{4}}+Q%
\left[ 20Q^{2}\left( 2Q^{2}-3r_{0}^{2}\right) +r_{0}^{4}\left(
r_{0}^{2}\left( 2r_{0}^{2}+45\right) -32Q^{2}\right) \right] \right\} }{%
45r_{0}^{2}\left( r_{0}^{2}-Q^{2}\right) ^{2}\left( 2\left(
r_{0}^{4}-Q^{2}\right) -Q\sqrt{4Q^{2}-2r_{0}^{4}}\right) }  \notag
\end{gather}%
which are only acceptable if $4Q^{2}-2r_{0}^{4}\geq 0.$ Specifically, once
the equality holds one has%
\begin{equation}
r_{0}^{4}=2Q^{2}
\end{equation}%
while from (16) we have 
\begin{equation}
r_{0}^{2}-2Mr_{0}+Q^{2}=0.
\end{equation}%
These together become%
\begin{equation}
r_{0}^{2}-2Mr_{0}+\frac{r_{0}^{4}}{2}=0
\end{equation}%
or%
\begin{equation}
M=r_{0}+\frac{1}{2}r_{0}^{3}.
\end{equation}%
Also in this case we find%
\begin{equation}
\beta _{1}=\frac{2}{3}\frac{Q}{r_{0}\left( r_{0}^{2}-Q^{2}\right) },
\end{equation}%
\begin{equation}
\beta _{2}=\frac{8}{45}\frac{4r_{0}^{2}-15}{\left( r_{0}^{2}-2\right) ^{2}}
\end{equation}%
and $f_{0}=R_{0}=1$ while $F_{0}=2.$

These examples can further be extended to cover more general polynomial
forms of $f(R)$ to justify the validity of our existence conditions,
however, we shall be satisfied with an extremal-RN example in the following
section.

\subsection{Extremal RN-type black hole}

An interesting case which can be considered here is the case for $M=Q$ in
(2). This will make the extremal RN-type black hole with the line element 
\begin{equation}
ds^{2}=-e^{-2\Phi }\left( 1-\frac{b_{0}}{r}\right) ^{2}dt^{2}+\frac{dr^{2}}{%
\left( 1-\frac{b_{0}}{r}\right) ^{2}}+r^{2}(d\theta ^{2}+\sin ^{2}\theta
d\varphi ^{2})
\end{equation}%
in which $b_{0}=Q.$ Taking this into account would lead from the general
equations 
\begin{equation}
R=6\frac{\beta }{r_{0}^{2}}\epsilon -\frac{6\beta }{r_{0}^{2}}\left( 2\beta +%
\frac{5}{r_{0}}\right) \epsilon ^{2}+\frac{\beta }{r_{0}^{2}}\left( \frac{%
93\beta ^{2}}{4}+\frac{71\beta }{r_{0}}+\frac{90}{r_{0}^{2}}\right) \epsilon
^{3}+O\left( \epsilon ^{4}\right) ,
\end{equation}%
\begin{equation}
f=6\frac{\beta }{r_{0}^{2}}\epsilon -\frac{3\beta }{r_{0}^{2}}\left( 3\beta +%
\frac{10}{r_{0}}\right) \epsilon ^{2}+\frac{\beta }{r_{0}^{2}}\left( \frac{%
57\beta ^{2}}{4}+\frac{49\beta }{r_{0}}+\frac{90}{r_{0}^{2}}\right) \epsilon
^{3}+O\left( \epsilon ^{4}\right) ,
\end{equation}%
\begin{equation}
F=\left( \frac{df}{dR}\right) =1+\beta \epsilon -\frac{\beta }{2}\left(
\beta +\frac{2}{r_{0}}\right) \epsilon ^{2}+\beta \left( \frac{3\beta ^{2}}{8%
}+\frac{3\beta }{4r_{0}}+\frac{1}{r_{0}^{2}}\right) \epsilon ^{3}+O\left(
\epsilon ^{4}\right)
\end{equation}%
and 
\begin{equation}
\Phi =\Phi _{0}+\beta \epsilon -\frac{\beta }{8}\left( 5\beta +\frac{8}{r_{0}%
}\right) \epsilon ^{2}+\beta \left( \frac{73\beta ^{2}}{120}+\frac{73\beta }{%
60r_{0}}+\frac{1}{r_{0}^{2}}\right) \epsilon ^{3}+O\left( \epsilon
^{4}\right)
\end{equation}%
in which $\beta $ is an arbitrary, non-zero constant and $\epsilon =\left(
r-r_{0}\right) .$ As before, we absorb $\Phi _{0}$ into time. It is
remarkable observe that $R$ is zero at the horizon and so is $f,$ but $%
\left( \frac{df}{dR}\right) =1.$ This is an indication that a proper
candidate for such an $f(R)$ is of the form%
\begin{equation}
f\left( R\right) =R+a_{2}R^{2}+a_{3}R^{3}+a_{4}R^{4}+...
\end{equation}%
in which the constant coefficients $a_{i}$ can be determined, using above
conditions. For instance, if we restrict ourselves up to the third order we
get $f\left( R\right) \sim R+\frac{r_{0}^{2}}{12}R^{2}+r_{0}^{3}\left( \frac{%
5}{72}r_{0}+\frac{19}{108\beta }\right) R^{3}$. One can easily check that
this form of $f(R)$ satisfies all the conditions given above up to the
second order. Subsequent implication of the results found above is that $%
f\left( R\right) \sim R^{\nu }.$ Here any $\nu $ can not satisfy the
conditions without choosing $\beta =0,$ which is the case of $\nu =1$ or $%
GR. $ Another example which at least satisfies the above conditions up to
first order is $f\left( R\right) =\frac{R}{1-R}.$

\section{Thermodynamics of the analog black hole}

After having the solution one may be curious about the thermodynamical
properties of the solution. This is doable in exact form because of the
metric function which is known about the horizon. First of all the horizon
will remain as $r=r_{0}$ and the Hawking temperature is found by%
\begin{equation}
T_{H}=\left. \frac{\frac{\partial }{\partial r}g_{tt}}{4\pi }\right\vert
_{r=r_{0}}=T_{H}^{\left( RN\right) }=\frac{1}{4\pi r_{0}}\left( 1-\frac{Q^{2}%
}{r_{0}^{2}}\right)
\end{equation}%
in which $T_{H}^{\left( RN\right) }$ implies RN Hawking temperature. The
form of Entropy is given by%
\begin{equation}
S=\left. \frac{\mathcal{A}}{4G}F\right\vert _{r=r_{0}}=\pi r_{0}^{2}F_{0}
\end{equation}%
in which $\left. \mathcal{A}\right\vert _{r=r_{0}}=4\pi r_{0}^{2}$ is the
surface area of the black hole at the horizon and $\left. F\right\vert
_{r=r_{0}}=F_{0}$. We note that $T_{H}$ and $S$ are both exact. Having $%
T_{H} $ and $S$ one may find the heat capacity of the black hole%
\begin{equation}
C_{q}=T\left( \frac{\partial S}{\partial T}\right) _{Q}=C_{q}^{(RN)}\mathcal{%
I}
\end{equation}%
in which 
\begin{equation}
\mathcal{I}=12Q^{2}\left( r_{0}^{2}-Q^{2}\right) \Pi
\end{equation}%
where%
\begin{equation}
\Pi =\frac{\left[ 5r_{0}^{3}\left( \left( r_{0}^{4}-Q^{4}\right) \beta
_{1}-4Q^{2}r_{0}\right) \beta _{2}+16r_{0}^{3}\beta _{1}^{3}\left(
r_{0}^{4}-Q^{4}\right) +4Q^{2}r_{0}^{2}\beta _{1}^{2}\left(
7r_{0}^{2}-23Q^{2}\right) +\frac{2Q^{2}\left( 24Q^{4}-r_{0}\beta _{1}\left(
15r_{0}^{4}+32r_{0}^{2}Q^{2}-59Q^{4}\right) \right) }{\left(
r_{0}^{2}-Q^{2}\right) }\right] }{\left[ r_{0}^{2}\left(
r_{0}^{2}-Q^{2}\right) ^{2}\left( 5\beta _{2}+16\beta _{1}^{2}\right)
+2r_{0}\left( r_{0}^{2}-Q^{2}\right) \left( 5r_{0}^{2}-23Q^{2}\right) \beta
_{1}+24Q^{4}\right] ^{2}}
\end{equation}%
We comment that $\mathcal{I}$ in the RN limit (i.e., $\beta _{i}\rightarrow
0 $) becomes unit as expected. Let us note also that the form of $C_{q}$ is
exact. In order to study the thermodynamics of the extremal solution we use
the general results found in the non-extremal RN type solution. One easily
finds that $T_{H}=0,$ and $C_{q}=0.$

\subsection{First Law of Thermodynamics}

Furthermore, in this section, we would like to show that in general, the
above solution also satisfies the first law of thermodynamics. This is
somehow a generalization of what was introduced in Ref. \cite{20} to find a
higher dimensional form of the Misner-Sharp (MS) energy \cite{21} and was
used in SSS black hole in $f(R)$ gravity in Ref. \cite{22}. To this end we
rewrite the field equation in the following form%
\begin{equation}
G_{\mu }^{\nu }=\kappa \left[ \frac{1}{F}T_{\mu }^{\nu }+\frac{1}{\kappa }%
\check{T}_{\mu }^{\nu }\right] ,
\end{equation}%
in which $G_{\mu }^{\nu }$ is the Einstein tensor, ~%
\begin{equation}
\check{T}_{\mu }^{\nu }=\frac{1}{f_{R}}\left[ \nabla ^{\nu }\nabla _{\mu
}F-\left( \square F-\frac{1}{2}f+\frac{1}{2}RF\right) \delta _{\mu }^{\nu }%
\right] 
\end{equation}%
and for our later convenience we consider 
\begin{equation}
ds^{2}=-e^{-2\Phi }Ud^{2}t+\frac{1}{U}d^{2}r+r^{2}d\Omega ^{2}.
\end{equation}%
In turn, the $tt$ component of the latter field equation would read%
\begin{equation}
G_{0}^{0}=\kappa \left[ \frac{1}{F}T_{0}^{0}+\frac{1}{\kappa }\frac{1}{F}%
\left[ \nabla ^{0}\nabla _{0}F-\left( \square F-\frac{1}{2}f+\frac{1}{2}%
RF\right) \right] \right] 
\end{equation}%
in which

\begin{equation}
G_{0}^{0}=\frac{U^{\prime }r-1+U}{r^{2}},
\end{equation}%
\begin{equation}
\nabla ^{0}\nabla _{0}F=\frac{1}{2}\left( -2\Phi ^{\prime }U+U^{\prime
}\right) F^{\prime }
\end{equation}%
and $\square F=\frac{2}{3}f-\frac{1}{3}RF.$ At the horizon (where the MS
energy is introduced) $U\left( r_{0}\right) =0,$ which yields $G_{0}^{0}=%
\frac{U_{0}^{\prime }r_{0}-1}{r_{0}^{2}},$ $\nabla ^{0}\nabla _{0}F=\frac{1}{%
2}U_{0}^{\prime }F_{0}^{\prime }.$ A substitution in (40) and calculating
everything at the horizon $r=r_{0}$ yields%
\begin{equation}
\frac{F_{0}U_{0}^{\prime }}{r_{0}}-\frac{F_{0}}{r_{0}^{2}}=\kappa
T_{0}^{0}+\left( \frac{1}{2}U_{0}^{\prime }F_{0}^{\prime }-\frac{1}{6}\left(
f_{0}+R_{0}F_{0}\right) \right) .
\end{equation}%
Next, we multiply both sides by the spherical volume element at the horizon
i.e. $dV_{0}=\mathcal{A}dr_{0}$ to get

\begin{equation}
\frac{F_{0}U_{0}^{\prime }}{r_{0}}\mathcal{A}dr_{0}=\left( \frac{F_{0}}{%
r_{0}^{2}}+\frac{1}{2}U_{0}^{\prime }F_{0}^{\prime }-\frac{1}{6}\left(
f_{0}+R_{0}F_{0}\right) \right) \mathcal{A}dr_{0}+\kappa T_{0}^{0}dV_{0}.
\end{equation}%
Using $\frac{\mathcal{A}}{r_{0}}=\frac{1}{2}\frac{d}{dr_{0}}\mathcal{A}$ and
some manipulation one finds

\begin{equation}
\frac{U_{0}^{\prime }}{4\pi }\frac{d}{dr_{0}}\left( \frac{2\pi \mathcal{A}}{%
\kappa }F_{0}\right) dr_{0}=\frac{1}{\kappa }\left( \frac{F_{0}}{r_{0}^{2}}%
+U_{0}^{\prime }F_{0}^{\prime }-\frac{1}{6}\left( f_{0}+R_{0}F_{0}\right)
\right) \mathcal{A}dr_{0}+T_{0}^{0}dV_{0}
\end{equation}%
which is nothing but the first law of thermodynamics i.e., $TdS=dE+PdV$.
This is due to the definition which we have for Hawking temperature $T=\frac{%
U_{0}^{\prime }}{4\pi },$ entropy of the black hole $S=\frac{2\pi \mathcal{A}%
}{\kappa }F_{0}$, the radial pressure $P=T_{r}^{r}=T_{0}^{0}$ and the MS
energy as 
\begin{equation}
E=\frac{1}{\kappa }\int \left( \frac{F_{0}}{r_{0}^{2}}+U_{0}^{\prime
}F_{0}^{\prime }-\frac{1}{6}\left( f_{0}+R_{0}F_{0}\right) \right) \mathcal{A%
}dr_{0}
\end{equation}%
in which the integration constant is set to zero \cite{20,23,24,25,26} (also
for a BH-like solutions see \cite{27}). Here we comment that all quantities
are calculated at the horizon and due to this the Hawking temperature
becomes $T=\left. \frac{\left( e^{-2\Phi }U\right) ^{\prime }}{4\pi }%
\right\vert _{r_{0}}=\frac{U_{0}^{\prime }}{4\pi }.$ The above results imply
that, using (65) as MS energy, the first law of thermodynamic is satisfied.
Once more we wish to add that our results are exact.

\section{Concluding Remarks}

In this paper we have applied the "\textit{near-horizon test}" to the
Reissner-Nordstr\"{o}m (RN)-type black holes in \ $f(R)$ gravity. Necessary
conditions, not the sufficient ones that a RN-type black hole exists are
derived. These are nothing but the regularity conditions of the metric
functions in the vicinity of the event horizon. Our metric ansatz consists
of a general static, spherically symmetric (SSS) case adopted from the
Einstein's general relativity. We considered also the extremal case as an
analog black hole in \ $f(R)$ gravity and derived the underlying conditions.
Due to their intricacy we didn't attempt to solve those equations in
general. To the zeroth order, however, they can be obtained exactly while to
the first order approximation is also tractable. Our analysis shows that a
closed form of \ $f(R)$ doesn't seem possible: With a given source we can
determine \ $f(R)$ implicitly as an infinite series in $\left(
r-r_{0}\right) ,$ since $R\left( r\right) $ also is expressed in similar
series. This is against the strategy adapted so far, namely, an explicit
form of \ $f(R)$ is assumed a priori to be tested whether it fits physical
requirements. In our opinion, the "\textit{near-horizon test}", introduced
in \cite{11,12} and developed here further constitutes a more fundamental
test than any other arguments in connection with black holes. We admit that
since our necessary conditions for the existence of RN-type black holes are
entirely local they don't involve the requirements for asymptotic flatness.
Stability of such black holes must also be considered separately when one
considers exact solutions. Our test must naturally be supplemented with $%
\frac{d^{2}f}{dR^{2}}>0$ and $\frac{df}{dR}>0$, for stability and no-ghost
requirements \cite{28,29}. We have shown also that the thermodynamic of
these analog black holes can be studied through the Misner-Sharp formalism
to verify the validity of the first law. Finally, we remark that solution
for $f(R)$ gravity admitting an electromagnetic field with similar
thermodynamics was reported before \cite{30}.


\begin{thebibliography}{99}
\bibitem{1} S. Nojiri and S. D. Odintsov, Phys. Rev. D \textbf{74}, 086005
(2006).

\bibitem{2} A. Sheykhi, Phys. Rev. D \textbf{86}, 024013 (2012).

\bibitem{3} M. Cvetic, S. Nojiri and S. D. Odintsov, Nucl. Phys. B \textbf{%
628}, 295 (2002).

\bibitem{4} R. G. Cai, Phys. Rev. D \textbf{65}, 084014 (2002).

\bibitem{5} A. de la Cruz-Dombriz, A. Dobado and A. L. Maroto, Phys. Rev. D 
\textbf{80}, 124011 (2009).

\bibitem{6} J. A. R. Cembranos, A. de la Cruz-Dombriz and P. Jimeno Romero,
"Kerr-Newman black holes in f(R) theories"arXiv:1109.4519.

\bibitem{7} S. Nojiri and S. D. Odintsov, Int. J. Geom. Methods Mod. Phys. 
\textbf{4}, 115 (2007).

\bibitem{8} T. P. Sotiriou and V. Faraoni, Rev. Mod. Phys. \textbf{82}, 451
(2010).

\bibitem{9} S. Nojiri and S. D. Odintsov, Phys. Rep. \textbf{505, }59 (2011).

\bibitem{10} L. Sebastiani, S. Zerbini, Eur. Phys. J. C \textbf{71}, 1591
(2011).

\bibitem{11} S. E. P. Bergliaffa and Y. E. C. de O. Nunes, Phys. Rev. D 
\textbf{84}, 084006 (2011).

\bibitem{12} S. H. Mazharimousavi and M. Halilsoy, Phys. Rev. D \textbf{86},
088501 (2012).

\bibitem{13} S. G. Ghosh and S. D. Maharaj, Phys. Rev. D \textbf{85}, 124064
(2012).

\bibitem{14} G. J. Olmo and D. Rubiera-Garcia, Phys. Rev. D \textbf{84},
124059 (2011).

\bibitem{15} A. Larranaga, Pramana Jou. of Phys. \textbf{78}, 697 (2012).

\bibitem{16} S. H. Hendi, B. Eslam Panah and S. M. Mousavi, Gen. Relativ.
Gravit. \textbf{44}, 835 (2012).

\bibitem{17} J. L. Said and K. Z. Adami, Phys. Rev. D \textbf{83}, 043008
(2011).

\bibitem{18} S. G. Ghosh and S. D. Maharaj, "Radiating Kerr-Newman black
hole in $f(R)$ gravity" arXiv:1208.3028.

\bibitem{19} G. J. Olmo1 and D. Rubiera-Garcia, "Charged black holes in
Palatini $f(R)$ theories" arXiv:1301.2091.

\bibitem{20} M. Akbar and R. G. Cai, Phys. Lett. B \textbf{648}, 243 (2007).

\bibitem{21} C. W. Misner and D. H. Sharp, Phys. Rev. \textbf{136}, B571
(1964).

\bibitem{22} S. H. Mazharimousavi and M. Halilsoy, Phys. Rev. D \textbf{84},
064032 (2011).

\bibitem{23} R. G. Cai, L. M. Cao, Y. P. Hu and N. Ohta, Phys. Rev. D 
\textbf{80}, 104016 (2009).

\bibitem{24} H. Maeda and M. Nozawa, Phys. Rev. D \textbf{77}, 064031 (2008).

\bibitem{25} M. Akbar and R. G. Cai, Phys. Rev.D \textbf{75}, 084003 (2007).

\bibitem{26} M. Akbar and R. G. Cai, Phys. Lett. B \textbf{635}, 7 (2006).

\bibitem{27} R. G. Cai, L. M. Cao, and N. Ohta, Phys. Rev. D \textbf{81},
084012 (2010).

\bibitem{28} I. L. Buchbinder, S. D. Odintsov and I. L. Shapiro, Effective
Actions in Quantum Gravity (IOP Publishing, Bristol, 1992).

\bibitem{29} G. A. Vilkovisky, Classical Quantum Gravity \textbf{9}, 895
(1992).

\bibitem{30} S. H. Mazharimousavi, M. Halilsoy and T. Tahamtan, Eur. Phys.
J. C. \textbf{72}, 1851 (2012).
\end{thebibliography}
\end{document}